\documentclass[journal=jpcl,manuscript=letter,layout=twocolumn]{achemso}
\usepackage{chemformula} % Formula subscripts using \ch{}
\usepackage[T1]{fontenc} % Use modern font encodings
\usepackage{braket}

\author{Yifan Quan}
\affiliation[MIT]
{Francis Bitter Magnet Laboratory and Department of Chemistry, Massachusetts Institute of Technology, Cambridge, Massachusetts 02139, United States}
\author{Jakob Steiner}
\affiliation[PSI]{Paul Scherrer Institute, 5232 Villigen PSI, Switzerland}
\author{Yifu Ouyang}
\affiliation[MIT]
{Francis Bitter Magnet Laboratory and Department of Chemistry, Massachusetts Institute of Technology, Cambridge, Massachusetts 02139, United States}
\author{Kong Ooi Tan}
\affiliation[MIT]
{Francis Bitter Magnet Laboratory and Department of Chemistry, Massachusetts Institute of Technology, Cambridge, Massachusetts 02139, United States}
\altaffiliation{Currently at Laboratoire des Biomol\'{e}cules, LBM, D\'{e}partement de Chimie, \'{E}cole Normale Sup\'{e}rieure, PSL University, Sorbonne Universit\'{e}, CNRS, 75005 Paris, France}
\author{W. Thomas Wenckebach}
\affiliation[PSI]{Paul Scherrer Institute, 5232 Villigen PSI, Switzerland}
\alsoaffiliation[MagLab]{National High Magnetic Field Laboratory, University of Florida, Gainesville, FL, United States}
\author{Patrick Hautle}
\affiliation[PSI]{Paul Scherrer Institute, 5232 Villigen PSI, Switzerland}
\author{Robert G. Griffin}
\affiliation[MIT]{Francis Bitter Magnet Laboratory and Department of Chemistry, Massachusetts Institute of Technology, Cambridge, Massachusetts 02139, United States}\email{rgg@mit.edu}

\title[ISE]
  {Integrated, stretched and adiabatic solid effects}

\abbreviations{IR,NMR,UV}
\keywords{Dynamic nuclear polarization, pulsed DNP, integrated solid effect, stretched solid effect, adiabatic solid effect}

\begin{document}


\begin{tocentry}

\includegraphics[width=5cm]{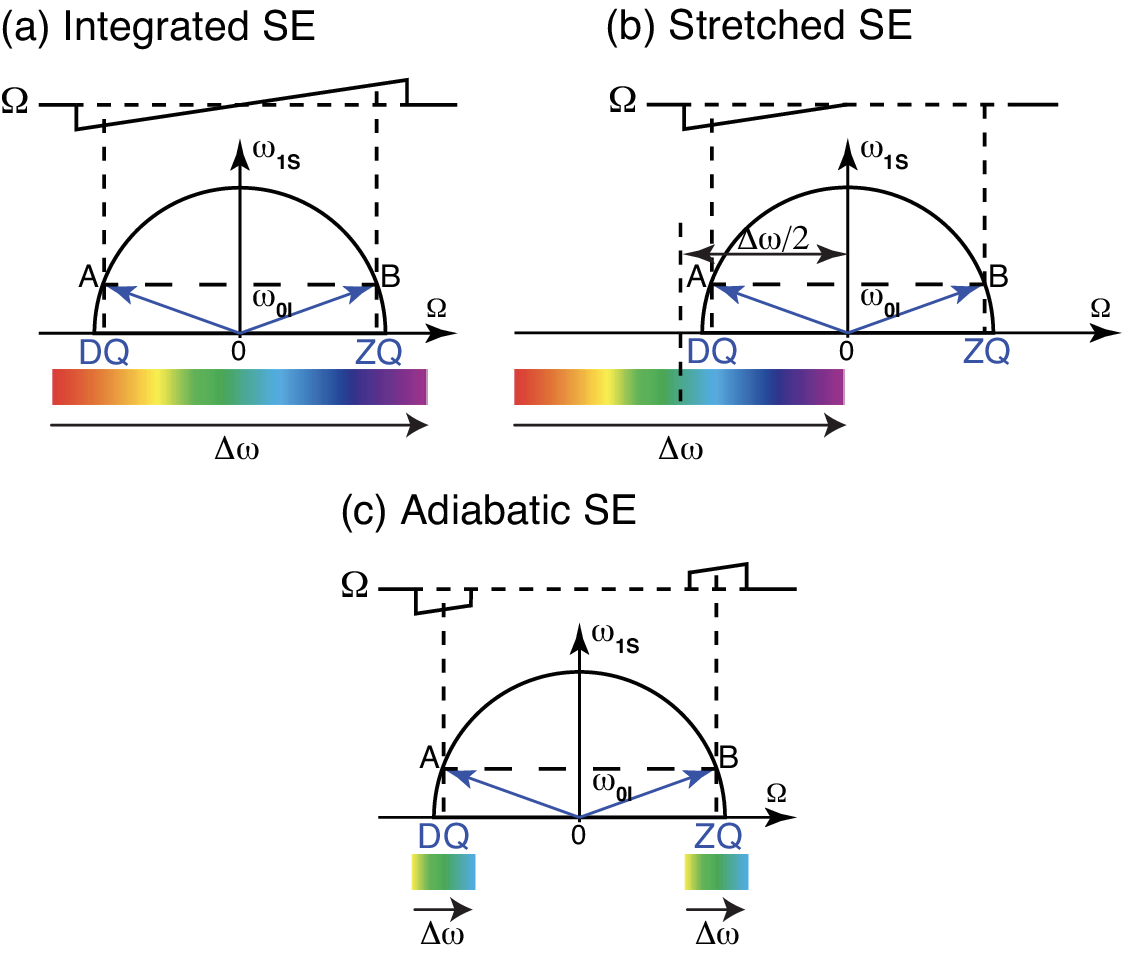}

\end{tocentry}

\begin{abstract}
\textbf{
This paper presents a theory describing the dynamic nuclear polarization (DNP) process associated with an arbitrary frequency swept microwave pulse. The theory is utilized to explain the integrated solid effect (ISE) as well as the newly discovered stretched solid effect (SSE) and adiabatic solid effect (ASE). It is verified with experiments performed at 9.4\,GHz (0.34\,T) on single crystals of naphthalene doped with pentacene-$d_{14}$.  It is shown that SSE and ASE can be more efficient than the ISE. Furthermore, the theory predicts that the efficiency of the SSE improves at high magnetic fields, where the EPR line width is small compared to the nuclear Larmor frequency. In addition, we show that ISE, SSE and ASE are based on similar physical principles and we suggest definitions to distinguish among them.}
\end{abstract}


\vspace{1cm}

The sensitivity of magnetic resonance experiments is the factor that chronically limits the success of any application. As a consequence, each advance in methodology and technology that has improved signal-to-noise in nuclear magnetic resonance (NMR), electron paramagnetic resonance (EPR) and magnetic resonance imaging (MRI) has erased the existing scientific boundaries, and initiated new areas of application and directions of research. A prominent example is dynamic nuclear polarization (DNP) that, in the last decade, has evolved in several guises as the method of choice for NMR signal enhancement.

However, in what are now routine continuous wave (CW) DNP experiments - the solid effect (SE) or cross effect (CE) - the signal enhancements, $\varepsilon$, exhibit a frequency dependence $\sim\omega_{0I}^{-1}$ to $\sim\omega_{0I}^{-2}$. Therefore, when performing experiments at the high magnetic fields, which are essential to optimize NMR  spectral resolution, signal enhancements drop significantly. In contrast, time domain or pulsed DNP mechanisms should not display an $\omega_{0I}$ dependence and thus provide a promising approach to enhance sensitivity in acquisition of high field NMR spectra.

One of the most favorable time domain DNP methods is the integrated solid effect (ISE), which was initially performed by sweeping the magnetic field through the SE matching conditions in a time $\sim T_{1e}$ \cite{Henstra:1988,Henstra:2014}. When the sweep is adiabatic, the polarization is coherently transferred from the electron to the surrounding nuclei. However, at high magnetic fields (9-28\,T) where most NMR is currently performed, it is technically challenging to rapidly sweep the Zeeman field, $B_0$. Alternatively, the Q of the microwave circuit is small, and therefore, the microwave frequency can be modulated with an arbitrary waveform generator (AWG) while keeping the magnetic field constant \cite{Can:2017b,Can:2018,DelageLaurin:2021}. Using this approach, it has been shown that ISE can be as efficient as nuclear orientation via electron spin locking (NOVEL) \cite{Henstra:2008,Can:2015}, the gold standard DNP mechanism.  Concurrently, the ISE requires much less microwave power than NOVEL\cite{Can:2018}, making it a promising method for high field experiments.

\begin{figure*}[h]
\centering
\includegraphics[width=2\columnwidth]{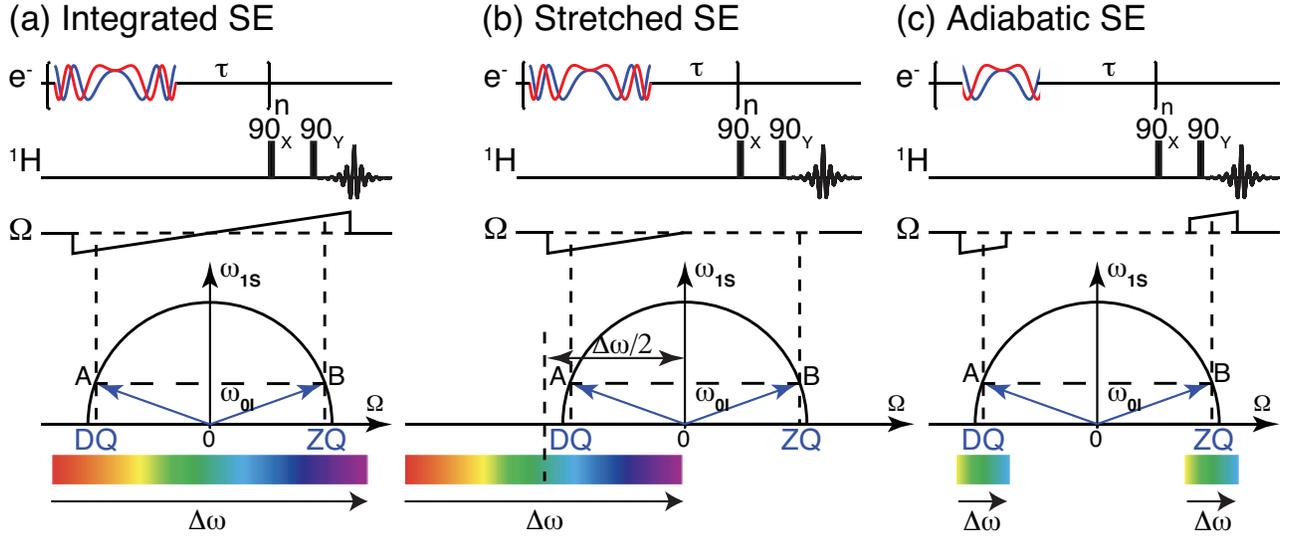}
\caption{Timing diagrams of the (a) ISE, (b) SSE and (c) ASE in the rotating frame. (a) ISE utilizes a broad microwave frequency sweep through both double quantum (DQ) and zero quantum (ZQ) matching conditions, e.g. $-\infty<\Omega<\infty$. (b) the SSE also uses a broad sweep but  only through one of the matching conditions, e.g. $-\infty<\Omega<0$. (c) the ASE uses a narrow sweep around one of the SE matching conditions at $\omega_{0S}\pm\omega_{0I}$. The EPR line width is not reflected in the schematic drawings.}
\label{fig:scheme}
\end{figure*}

In addition to the ISE, two closely related chirped pulse experiments were recently developed.  First, in performing experiments where the chirped frequency is not centered on the EPR spectrum, we observed a new polarization mechanism the stretched solid effect (SSE) \cite{Can:2018}. Second, when the chirped pulse is swept across one of the solid effect (SE) transitions at $\omega_{0S}\pm\omega_{0I}$  the signals were enhanced by a factor of $\sim 2.4$ over the SE enhancement, an effect which we term an adiabatic solid effect (ASE) \cite{Tan:2020}. Figure \ref{fig:scheme} illustrates the frequency modulation schemes of all three of these experiments -- the ISE, SSE and ASE. Furthermore, although it was observed experimentally that the SSE and ASE can be more efficient than a full ISE,  a theoretical understanding of this observation and these two DNP mechanisms is not yet available. Finally, the intellectual boundaries between ISE, SSE and ASE are not well defined. This situation provides the motivation for the development of a comprehensive theory for arbitrary chirped pulse DNP which will clarify the polarization transfer mechanism of these three DNP methods and hopefully lead to their further improvement.

%In a realistic DNP sample, there are a large number of electrons interacting with an even larger number of nuclear spins. The electron spins are very diluted, thus the electron electron spin interaction is neglected and the system is broken down into systems consisting a single electron spin $S$ interacting with $N_I$ nuclear spins $I$, which is the spin system for the theoretical development. In addition the nuclear nuclear dipolar interaction is neglected since that is much weaker than the hyperfine coupling and not involved in the polarization transfer process. The nuclear nuclear dipolar interaction is involved in the spin diffusion, which is a secondary process after the polarization transfer but is absolutely critical for DNP to polarize the bulk nuclei.

To calculate the polarization transfer of an arbitrary chirped pulse, we employ Landau-Zener and single value decomposition theory that were used previously to describe the ISE \cite{Henstra:2014}. We consider a spin system consisting of a single electron spin $S$ interacting with $N_I$ nuclear spins $I$, as the electron spin is typically very dilute. In addition, we neglect the internuclear dipolar interaction since it is much weaker than the electron-nuclear dipole coupling and not involved in the direct polarization transfer process. The internuclear dipolar coupling is essential for spin diffusion, which is a secondary process after the polarization transfer and is required for the DNP process to polarize the bulk nuclei.
%The Hamiltonian in the rotating frame can be written as

%\begin{equation}
%\begin{aligned}
%\label{equ:Hamiltonian}
%{\cal H}^r&=(\omega_{0S}-\omega)S_z+\omega_{1S}S_x^r-\omega_{0I}\sum_i^{N_I}I_z^i+\frac{1}{2}\sum_i^{N_I}S_z(2A_{zz}^iI_z^i+A_{z+}^iI_-^i+A_{z-}^iI_+^i).
%\end{aligned}
%\end{equation}
%Here $\omega_{0S}$ and $\omega_{0I}$ are the electron and nuclear Larmor frequencies. $\omega$ denotes the microwave frequency, which can be written as $\omega=\dot\omega t$ with $\dot\omega$ denoting the frequency sweep speed of the chirp pulse, and $\omega_{1S}$ denotes the microwave Rabi frequency. $S_z$ and $S_x^r$ are the electron spin operators in the rotating frame. Furthermore, we introduce the step operators for the nuclear spins $I_\pm^i=I_x^i \pm iI_y^i$, and $A_{z\pm}^i=A_{zx}^i \pm iA_{zy}^i$ denotes the hyperfine interaction.

Assuming an adiabatic sweep from $-\infty$ and an initial nuclear polarization, $P_I=0$, the electron polarization after matching point $A$ and subsequently point $B$ are written down using equations (70) and (72) from Henstra et al. \cite{Henstra:2014}.
\begin{equation}
\label{equ:tran_1match}
P_S^A=\int_0^\infty d\sigma^2 g(\sigma^2) P_S^0(2{\cal P}-1)=P_S^0\frac{1-{\cal Q}}{1+{\cal Q}},
\end{equation}
and
\begin{equation}
\begin{aligned}
\label{equ:tran_2match}
P_S^B&=\int_0^\infty d\sigma^2 g(\sigma^2) P_S^0(2{\cal P}-1)^2\\
&=P_S^0(1- \frac{4}{(1+2{\cal Q})(1+{\cal Q}^{-1})}).
\end{aligned}
\end{equation}
Here, $P_S^0$ denotes the initial electron polarization, $\sigma$ the hyperfine interaction strength with a spectral density $g(\sigma^2)=\frac{1}{M_2}\textrm{exp}(-\frac{\sigma^2}{M_2})$ where $M_2$ is the second moment, and
\begin{equation}
\label{equ:p_value}
{\cal P} \; = \;
\exp \left( - \; 2\pi \;
\frac{\sin^2 \theta_{\rm A} \sigma^2}{
|\cos\theta_{\rm A} \dot{\omega}|} \right),
\end{equation}
where $\sin\theta_{\rm A}=\omega_{1S}/\omega_{0I}$ denotes the extent of state mixing and $\dot\omega$ is the rate of the frequency sweep. Furthermore
\begin{equation}
\label{equ:Q_value}
{\cal Q} \; = \;
2\pi
\frac{\sin^2 \theta_{\rm A}M_2}{
|\cos\theta_{\rm A} \dot{\omega}|}.
\end{equation}
Notice that {\cal P} and {\cal Q} are derived from the Landau-Zener theory and the definition of $\cal Q$ here is different by a factor of $\sqrt{2}$ from that defined by Henstra et. al. \cite{Henstra:2014}. In addition, the sweep needs to be adiabatic, i.e. $\frac{\pi\omega_{1S}^2}{2|\dot\omega|}\gg1$ otherwise the polarization is inefficient.

\begin{figure}[H]
\centering
\includegraphics[width=\columnwidth]{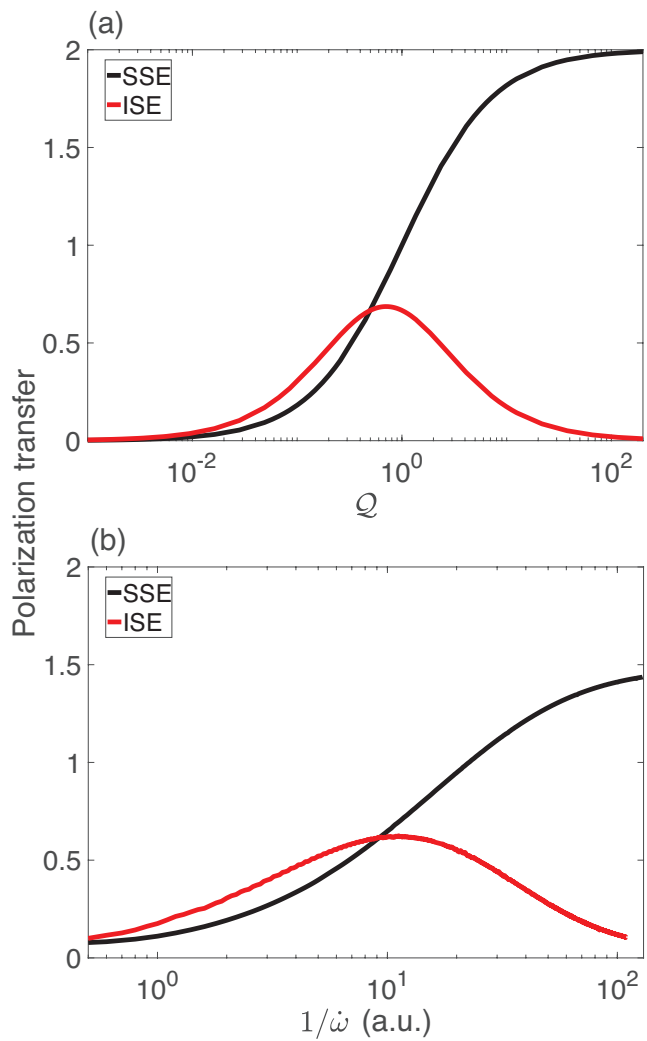}
\caption{(a) Electron polarization transfer efficiency for SSE and ISE calculated from (\ref{equ:tranA}) and (\ref{equ:tranB}) as well as (b) the simulation using the SpinEvolution software. For the latter, a system of one electron spin interacting with 9 nuclear spins is used, where the hyperfine interactions between the electron and each nuclear spin are set to be the same. The theoretical calculation is plotted as a function of ${\cal Q}$, while the SpinEvolution simulation plot is plotted as a function of $1/\dot\omega$ (see (\ref{equ:Q_value})) as there are only limited number (nine) of nuclear spins in the simulated system, hence no spectral density. The units for $1/\dot\omega$ in the simulation plot are arbitrary as they depend on the input hyperfine coupling strength. Both plots are with the horizontal axis on a logarithmic scale.}
\label{fig:1enH}
\end{figure}

From (\ref{equ:tran_1match}) and (\ref{equ:tran_2match}) we can write down the polarization transfer efficiency, $E_{\textrm{SSE}}$ and $E_{\textrm{ISE}}$, for a given nuclear polarization $P_I$,
\begin{equation}
\label{equ:tranA}
E_{\textrm{SSE}} \; = \; \frac{P_S^0 - P_S^A}{P_S^0-P_I} \; = \;
\frac{2{\cal Q}}{1 + {\cal Q}},
\end{equation}
and
\begin{equation}
\label{equ:tranB}
E_{\textrm{ISE}} \; = \; \frac{P_S^0 - P_S^B}{P_S^0-P_I} \; = \;
\frac{4}{\left( 1 + 2{\cal Q} \right)
\left( 1 + {\cal Q}^{-1} \right)}.
\end{equation}
These quantities are plotted in Figure \ref{fig:1enH} (left). It should be pointed out that the efficiency $E_{\textrm{ISE}}$ for a full ISE has been  studied in detail \cite{Henstra:2014,Eichhorn:2014,Quan:2019}, whereas the efficiency $E_{\textrm{SSE}}$ for sweeping through a single matching condition has not. Only recently, did we realize that sweeping through both transitions is not necessarily more efficient than sweeping only one, as shown in Figure \ref{fig:1enH} (left). This illustrates the underlying physics that leads to a higher efficiency for the SSE and ASE as opposed to the ISE. For the moment we define the ISE as a sweep through both SE matching conditions, whereas the SSE as through only one. From   (\ref{equ:Q_value}) we see that $0<{\cal Q}<\infty$ and the maximum of $E_{\textrm{SSE}}$ is 2 when ${\cal Q}\rightarrow\infty$, while that of $E_{\textrm{ISE}}$ is only 0.6863 at ${\cal Q}=1/\sqrt{2}$. ${\cal Q}\rightarrow\infty$ can be approached by an extremely slow sweep $\dot\omega\rightarrow0$ (see (\ref{equ:Q_value})) and $E_{\textrm{SSE}}=\frac{P_S^0 - P_S^A}{P_S^0-P_I}=2$  which means that $P^A=-P_S^0$ when $P_I^0=0$, i.e. a full inversion of the electron polarization with respect to the effective field. Thus, when subsequently going through the second matching point, the nuclei are polarized towards the opposite direction, reducing the nuclear polarization (polarization is transferred back from the surrounding nuclei to the electron).

The theory is first compared to numerical simulations performed using the SpinEvolution \cite{Veshtort:2016} software, which is based on calculating the evolution of the density matrix. The available computing power limits the number of spins that can be used in the calculation. We performed the simulation on a system of a single electron interacting with 9 nuclear spins, shown in Figure \ref{fig:1enH}, and compared it to the theoretical values given by (\ref{equ:tranA}) and (\ref{equ:tranB}). The hyperfine interactions between the electron and the 9 nuclei are set to be equal. The SpinEvolution simulation confirms the theoretical prediction that ISE has an optimum ${\cal Q}$ or $\dot\omega$, while SSE is always more efficient with higher ${\cal Q}$ or slower frequency sweep. Notice that the number of the nuclear spins in the simulated system is critical in studying DNP, since the electron polarization transfer efficiency increases when the number of nuclear spins increases. The maximum polarization transfer efficiency of the SSE can be higher than 1 for a system with many nuclear spins (as shown in Figure \ref{fig:1enH} for 9 nuclear spins). This means that the electron polarization can be inverted with respect to the effective field as theory predicts. With more nuclear spins in the system, the SSE polarization transfer efficiency is approaching the theoretical prediction of the maximum 2.

One of the reasons for preferring the ISE is that to date pulsed-DNP experiments have been limited to low magnetic fields, where the EPR line width is broader than the nuclear Larmor frequency. In this case, a sweep from $-\infty$ to 0 only brings part of the electron spins through one matching condition, part through both, and the rest through zero matching condition. We can treat this situation by considering the entire EPR line as spin packets with spectral density $f(\Omega)$ and the polarization transfer of each spin packet is calculated separately. A finite sweep from $\omega_i$ to $\omega_f$ instead of from $-\infty$ transfers only a fraction of the electron polarization that is parallel to the effective field to the nuclei instead of the full electron polarization. The fraction of the polarization parallel to the effective field is given by
\begin{equation}
\label{equ:P_frac}
%P^i_S(\Omega)= \pm P_S^0\cos(\arctan\frac{\omega_{1S}}{\omega_{0S}+\Omega-\omega_i}),
P^i_S(\Omega)= \pm P_S^0\frac{\omega_{0S}+\Omega-\omega_i}{\sqrt{\omega_{1S}^2+(\omega_{0S}+\Omega-\omega_i)^2}},
\end{equation}
where $\omega_{0S}+\Omega$ is the electron Larmor frequency of a spin packet at frequency offset $\Omega$ and the sign of $P_S$ depends on the sign of $\omega_{0S}+\Omega-\omega_i$.

Now we can write down the polarization transfer efficiency for any sweep and EPR line
\begin{equation}
\label{equ:Pf}
\begin{aligned}
E &= \int d\Omega f(\Omega) (P^i_S(\Omega)-P_S^f(\Omega)) \\
&= \int_A d\Omega f(\Omega) (P^i_S(\Omega)-P_I)E_{\textrm{SSE}}\\
&\ \ \ \ + \int_B f(\Omega) d\Omega (P^i_S(\Omega)-P_I) E_{\textrm{ISE}} \\
&= \int_A d\Omega f(\Omega) (P^i_S(\Omega)-P_I)\frac{2{\cal Q}}{1 + {\cal Q}} \\ &\ \ \ \ + \int_B d\Omega f(\Omega) (P^i_S(\Omega)-P_I) \frac{4}{\left( 1 + 2{\cal Q} \right)\left( 1 + {\cal Q}^{-1} \right)},
\end{aligned}
\end{equation}
where the interval $A$ and $B$ denote the spin packets going through a single or both matching points, respectively. Note that part of the spins that do not go through either of the matching points, induce no polarization transfer and contribute nothing to the integral. Here we have neglected the non-adiabatic process happening due to a sudden start and stop of the microwave pulse.

We immediately realize that ASE deploying a narrow chirp pulse around one of the SE line \cite{Kong:2020} is exactly the case that the chirp pulse brings a fraction of the polarization through one of the matching conditions. Now that we have a clearer understanding of the spin dynamics induced by an arbitrary chirp pulse, we can better separate ISE, SSE and ASE, defined as follows. Demonstrated in Figure \ref{fig:scheme}, an ISE is a broad sweep bringing all of the electron spins through both SE matching conditions, e.g. from $-\infty$ to $\infty$, while an SSE is a broad sweep bringing most of the electron spins through one of the matching conditions, e.g. from $-\infty$ to $\omega_{0S}$. Finally, the ASE is a narrow sweep around one of the SE lines.

\begin{figure}[H]
\centering
\includegraphics[width=\columnwidth]{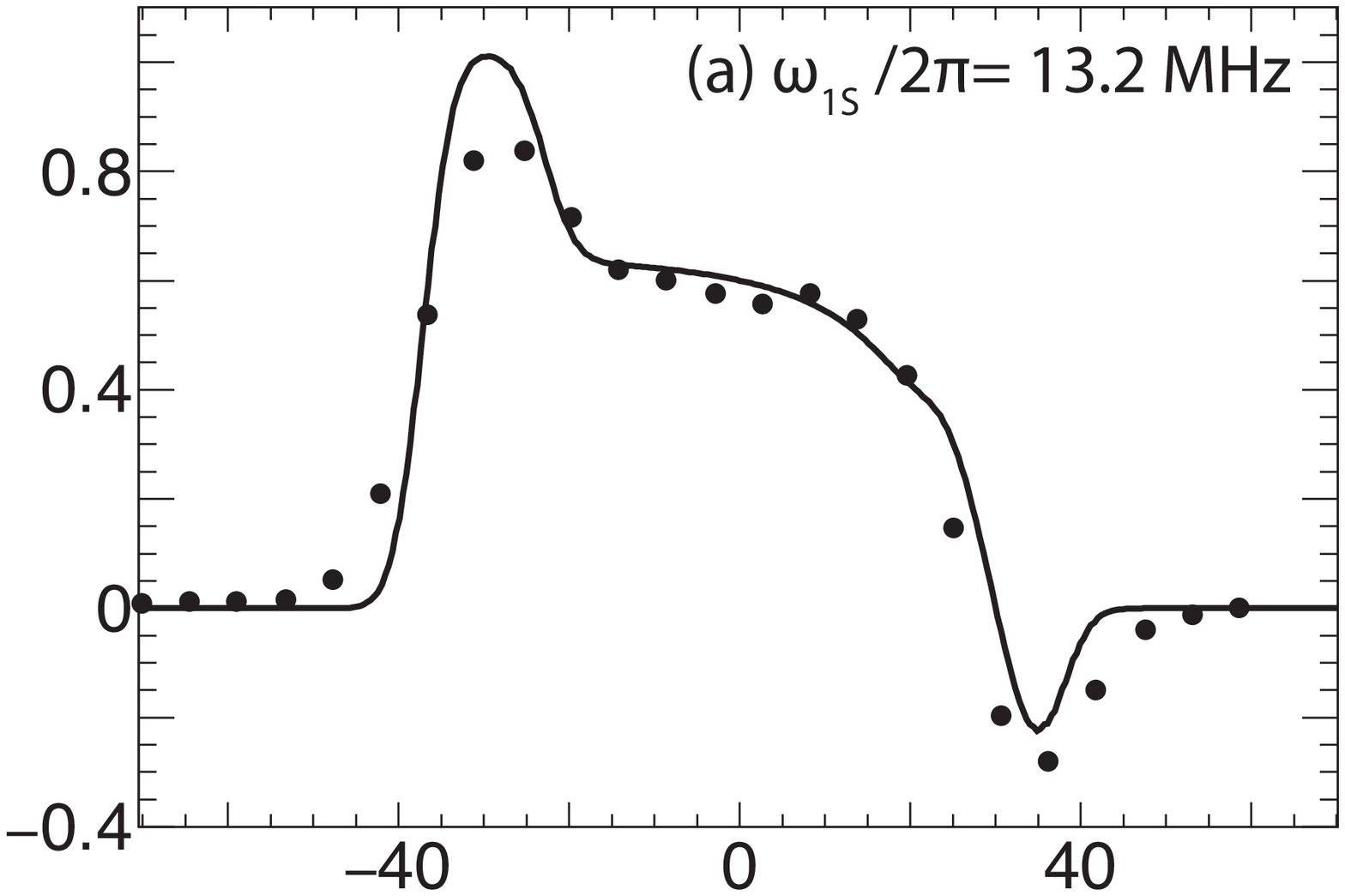}
\includegraphics[width=\columnwidth]{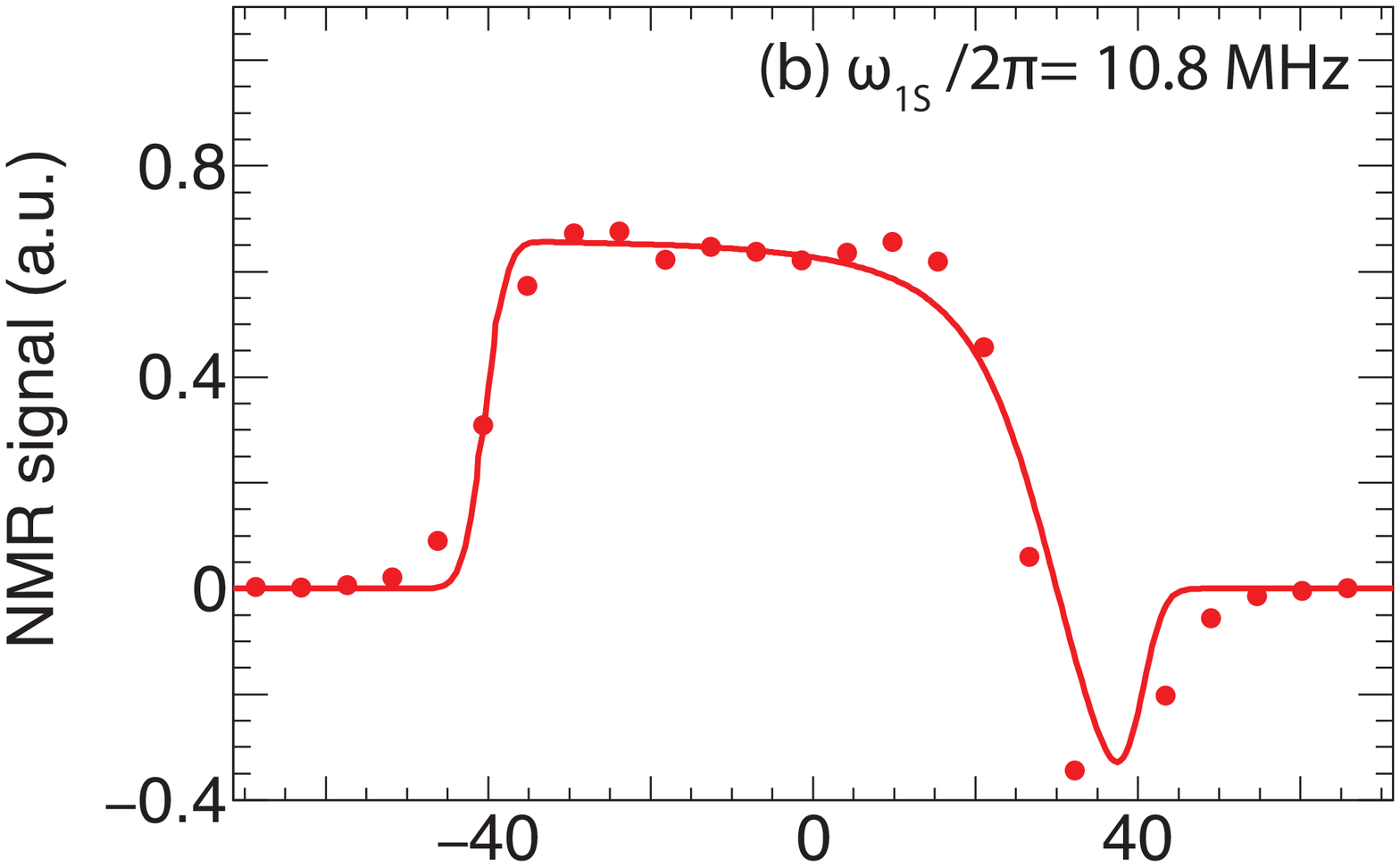}
\includegraphics[width=\columnwidth]{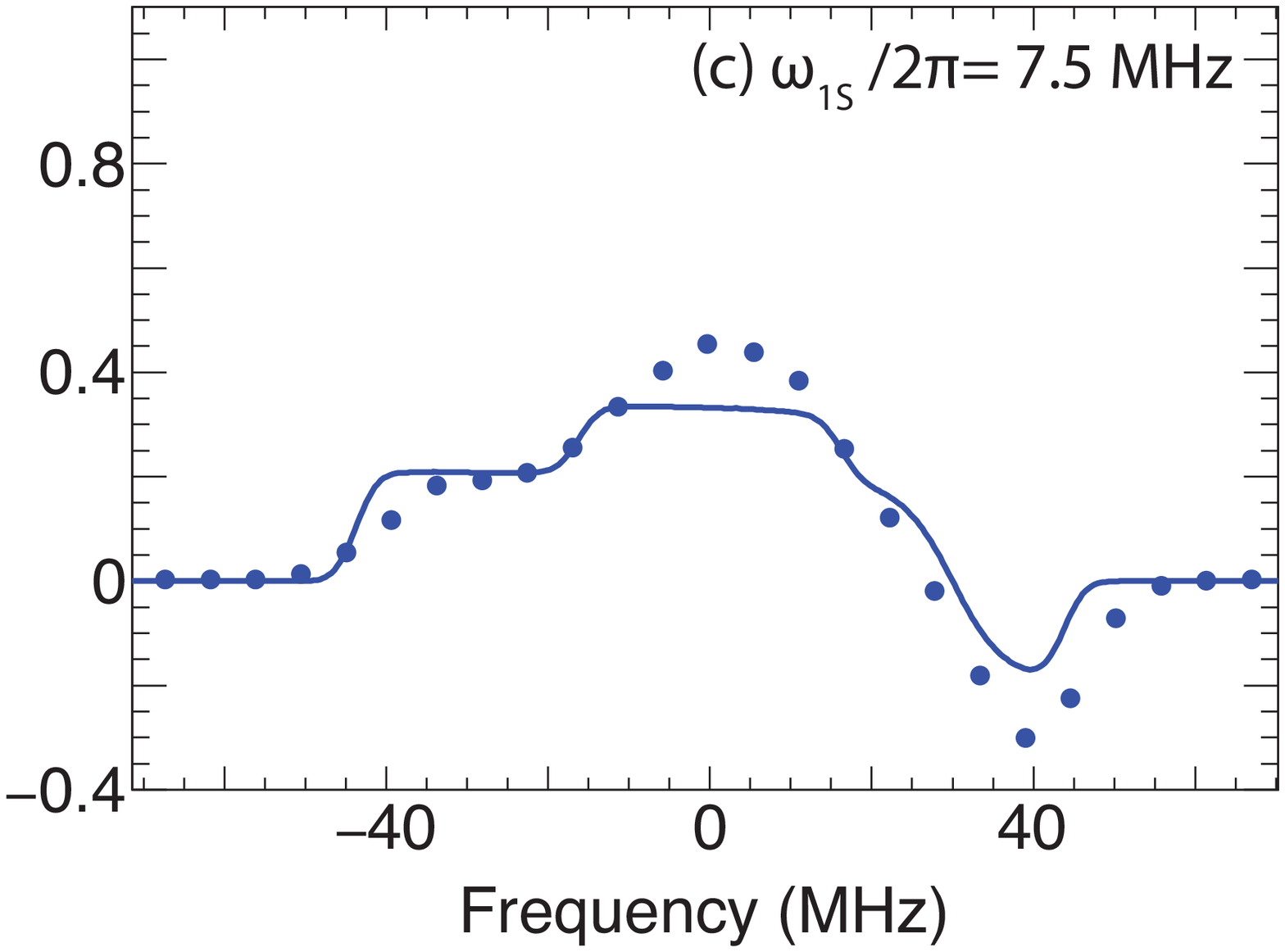}
\caption{Field profile of the proton polarization build-up rate in a naphthalene single crystal at T= 25 K, with different Rabi frequencies, $\omega_{1S}$. A  total sweep width of 22\,G ($\sim60$\,MHz) with $\dot\omega=1.5$\,G/$\mu$s (4.2\,MHz/$\mu$s) was employed and the horizontal axis indicates to the center of the sweep. They are fit to the theoretical model (Equation \ref{equ:Pf}), which are drawn as the solid lines.}
\label{fig:ISE_experiment}
\end{figure}

Shown in Figure \ref{fig:ISE_experiment} is the measured field profile of the proton polarization build-up rate in a naphthalene single crystal doped with pentacene-$d_{14}$ \cite{Eichhorn:2013} polarized by triplet-DNP \cite{Quan:2019} at X-band. The sweep width was  22\,G ($\sim60$\,MHz) with $\dot\omega=1.5$\,G/$\mu$s (=\,4.2\,MHz/$\mu$s) at three different Rabi fields $\omega_{1S}$ (details see below). The photo-excited triplets of the pentacene-$d_{14}$ guest molecules serve as the polarizing agent. The plateau around the center shows the ISE effect, while around -30\,MHz we find the SSE effect is maximal. This is fitted to the analytical calculation of (\ref{equ:Pf}), fixing all the known parameters. We assume that the EPR line to be a Gaussian distribution with a 3\,MHz standard deviation. We obtain an estimate of $\sqrt{M_2}/2\pi=0.27$\,MHz, compared to $\sqrt{M_2}/2\pi=0.5$\,MHz given by Eichhorn et. al \cite{Eichhorn:2014}. The discrepancy can be explained by the following considerations. Here, a much longer pulse is applied, leading to more triplets decaying into their ground states in the middle of each DNP pulse. In addition, the level crossing becomes non-linear (see (52) in Henstra et. al. \cite{Henstra:2014}) and can no longer be precisely described by the Landau-Zener theory. These can also explain the non-perfect fit to the measured DNP field profiles.

The experimental results are in excellent agreement with the our theoretical calculation. With a small Rabi frequency $\omega_{1S}$ (equivalent to a fast sweep, see equation (\ref{equ:Q_value})), the SSE is smaller than the ISE at the spectrum center. However, when $\omega_{1S}$ is large (equivalent to a slow sweep), then the SSE is more efficient than the ISE. The SSE is always more efficient when increasing $\omega_{1S}$ (or reducing sweep speed), while the ISE has an optimum $\omega_{1S}$ (or sweep speed). These features are predicted by the theory and embodied in equations (\ref{equ:tranA}), (\ref{equ:tranB}) and Figure \ref{fig:1enH}. Because the EPR line is symmetric, the frequency (field) with the optimal SSE efficiency is exactly a sweep to the center EPR frequency. For certain radicals, whose EPR spectrum is asymmetric due to a g-anisotropy, there could be an offset for the optimum frequency (field) depending on the detailed shape of the spectrum.

\begin{figure}[h]
\centering
\includegraphics[width=0.9\columnwidth]{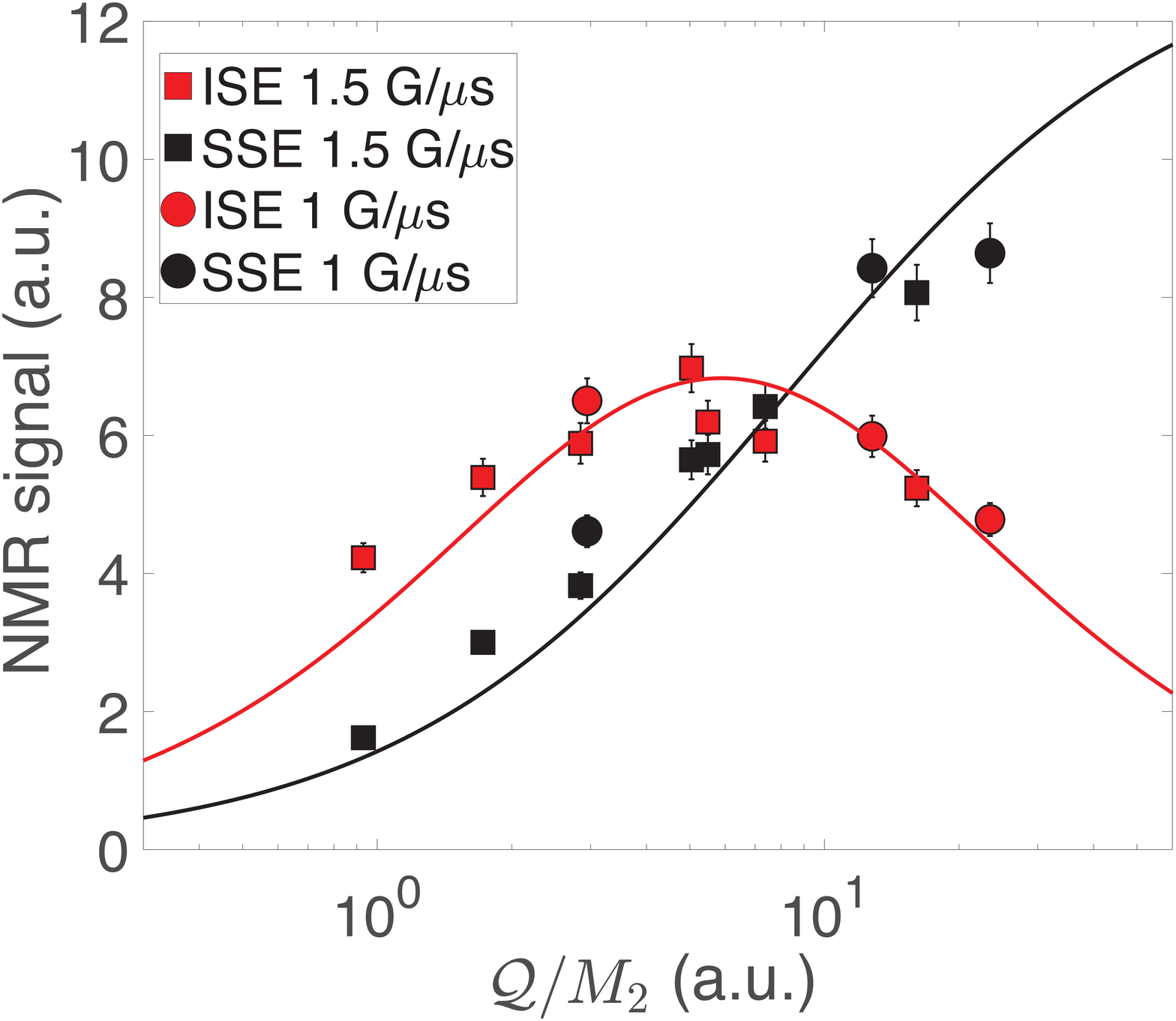}
\caption{Proton polarization build-up rate measured by NMR as a function of ${\cal Q}/M_2$ (see (\ref{equ:Q_value})) for ISE (0\,MHz in Figure \ref{fig:ISE_experiment}) and SSE (-30\,MHz in Figure \ref{fig:ISE_experiment}), where two sets of data with $\dot{\omega}=1.5$\,G/$\mu$s and $\dot{\omega}=1$\,G/$\mu$s are combined. The solid lines are drawn as guides to the data .}
\label{fig:ISE_Q}
\end{figure}

Figure \ref{fig:ISE_Q} summarizes the proton polarization build-up rate for ISE and SSE as a function of $\frac{\sin^2\theta}{\cos\theta\dot\omega}$ that is proportional to ${\cal Q}$ (equation (\ref{equ:Q_value})). Two sets of measurements with different sweep rates, $\dot{\omega}=1.5$\,G/$\mu$s (same as Figure \ref{fig:ISE_experiment}) and $\dot{\omega}=1$\,G/$\mu$s, are combined. This further confirms the theoretical prediction that the ISE has an optimum while the SSE keeps improving with higher Rabi fields (higher $\theta$) and slower sweep rates.

\begin{figure}[h]
\centering
\includegraphics[width=\columnwidth]{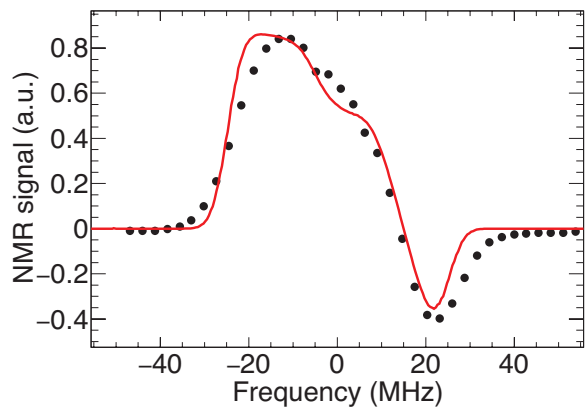}
\caption{DNP field profile of the $^1$H polarization build-up rate in a naphthalene single crystal at T=25 K measure by NMR with a 10\,G ($\sim28$\,MHz) field sweep (the sweep center is plotted as the horizontal axis) with $\dot\omega=1$\,G/$\mu$s (2.8\,MHz/$\mu$s). This experimental data is compared to the analytical calculation using equation (\ref{equ:Pf}) represented as the solid line.}
\label{fig:ASE_experiment}
\end{figure}

In addition, the field profile with a narrower sweep 10\,G (28\,MHz\,$\sim$\,EPR line width) with 1\,G/$\mu$s (=\,2.8\,MHz/$\mu$s) was recorded to demonstrate the theory to the ASE.  The data agrees well with the simulation as illustrated in Figure \ref{fig:ASE_experiment}.

To summarize, we have presented a theory for calculating the DNP polarization transfer for any chirped pulse on or near an arbitrarily shaped EPR line. The theory is based on the Landau-Zener equations, and we verified its primary features with DNP measurements of the $^1H$ polarization build-up rate for the ISE, SSE and ASE experiments.

Importantly, we showed that ISE, SSE and ASE are based on the same physical principles as the SE and NOVEL, where the energy levels of the electron and the surrounding nuclei are matched for mixing in the rotating frame. The ISE, SSE and ASE employ adiabatic frequency swept chirped pulses (or magnetic field sweeps), and therefore can utilize the electron polarization of the entire EPR line, which is an advantage over NOVEL. The ramped-amplitude NOVEL (RA-NOVEL) experiment \cite{Can:2017} also addresses this problem and yields improved enhancements. However, the ISE, SSE and ASE require much less microwave power than the NOVEL and RA-NOVEL experiments.

The calculations also permit us to define borders among the ISE, SSE and ASE. As we have seen the ISE employs a broad frequency sweep, e.g. from $-\infty$ to $+\infty$, where all of the electron spins transition through both matching conditions. The SSE also employs a  broad sweep that brings most of the electron spins through one of the matching conditions, e.g. from $-\infty$ to $\omega_{0S}$, the center of the EPR line.  In contrast, the ASE is a narrow sweep around one of the SE lines. In  an SSE or ASE, the majority of the electron spins transition through only one of the two matching points. Importantly, we have theoretically and experimentally demonstrated that, when the Rabi field $\omega_{1S}$ is strong and the sweep rate $\dot\omega$ is slow, then  going through the matching point once yields the optimum signal enhancement. This is because the electron polarization can be efficiently transferred to the nuclei (inverted in the extreme case) after going through the first matching condition.  A subsequent passage through the second matching condition actually transfers negative polarization from the electrons to the surrounding nuclei or transfers nuclear polarization to the electron spin as shown by (\ref{equ:tran_1match}) and (\ref{equ:tran_2match}) when $2{\cal P}-1<0$. This was confirmed using density matrix simulations and the program SpinEvolution.  Furthermore, a comparison of (\ref{equ:tranA}) and (\ref{equ:tranB}) shows that an optimal SSE can be a factor of $\sim3$ ($2/0.6863$) more effective than an optimal ISE. This theoretical maximum may be approached when the EPR line is  narrower than the nuclear Larmor frequency, a situation which exist in experiments at high magnetic fields when trityl or BDPA are the polarizing agents. At low magnetic fields, in a X-band system, we currently observed a $\sim$30\% enhancement or the SSE compared to ISE.

In addition, the electron polarization after an SSE is in the $xy$-plane perpendicular to the magnetic field, yielding a zero electron polarization. However, an ISE pulse brings the electron polarization from parallel to antiparallel to the magnetic field $z$-direction. Therefore, in SSE experiments, it requires a shorter period for the electron spin polarization of stable radicals to recover to the $+z$ direction when compared to ISE.  Thus, higher repetition rates and improved signal-to-noise is available from the SSE DNP experiment. One could consider returning the electron polarization to parallel to the magnetic field $z$-direction after each DNP (of any kind) pulse. This idea was suggested as a method to improve the execution of NOVEL \cite{Jain:2017}.

\section{Experimental methods}

The experimental studies to verify the theory have been performed with a well known triplet DNP system, a naphthalene molecular host crystal doped with a small amount of pentacene-$d_{14}$ guest molecules \cite{Eichhorn:2013CPL}. With a short laser pulse pentacene is excited into a triplet state, which is strongly aligned due to the selection rules of the photo-excitation process \cite{vanstrien:1980}. 
%The triplet $\ket{m_s} = \ket{0}$ state is highly populated with a relative occupation of $\rho_0 = 0.94$ and a lifetime of 73\,$\mu$s at the costs of the triplet  $\ket{m_s} = \ket{+1}$ and $\ket{-1}$ states with $\rho_- = \rho_+ = 0.03$ with lifetimes of 167\,$\mu$s \cite{Niketic:2017}.
%Application of a magnetic field along the molecular $X$-axis of the pentacene, leads to two ESR transitions between the eigenstates $\ket{m_s} = \ket{0}$ and $\ket{1}$ and $\ket{m_s} = \ket{0}$ and $\ket{-1}$ of the z-component $S_z$ that are separated by about 1.5\,GHz at 0.3\,T. For DNP the so called high field transition between $\ket{m_s} = \ket{0}$ and $\ket{1}$ is used, where the triplet spin lattice relaxation time below 100\,K is negligible \cite{Ong:1995}.

High quality crystals are grown from extensively zone refined naphthalene doped with custom synthesized pentacene-d$_{14}$ (ISO-TEC, Sigma Aldrich Group) by a self-seeding Bridgman technique \cite{Selvakumar:2005}. The pentacene concentration of the one used for the experiments was determined with optical transmission spectroscopy to be $(7.9\pm 0.5)\times 10^{-5}$\,mol/mol. 
%A rectangular sample was cut from this crystal measuring 3.6\,mm along the crystalline $a$-axis, 1.8\,mm along the crystalline $b$-axis and 4.2\,mm perpendicular to the $ab$-plane. It is mounted on a polychlorotrifluorethylene (PCTFE) holder, 
The sample was introduced in a homebuilt efficient helium flow cryostat sitting between the pole pieces of a compact electro-magnet. The cryostat is typically operated at 25\,K during the DNP process. The pentacene molecules are photo-excited using a 556\,nm laser (CNI HPL-556-Q 50) operating at 1\,kHz. 
%The light is transported from the laser system via a multimode fiber to an optical stage at the bottom of the cryostat collimating the unpolarized light through a sapphire optical window to a beam waist of about 9\,mm. The beam axis is vertical and along the b-axis illuminating the sample crystal homogeneously.

A versatile homebuilt pulse X-band ESR system, operating at 9.3 GHz and synchronized to the laser was assembled to perform pulse DNP experiments \cite{Eichhorn:2013}. 
%It also enables observation of the ESR signal of the triplet states and allows the crystal to be oriented so that the static magnetic field is along the $X$-axis of the pentacene molecules. The sample sits in the middle of a high-Q resonator, consisting of a sapphire dielectric ring placed inside a cylindrical TE011 brass cavity. 
A 10 turn saddle coil wound on a teflon frame around the cavity provides the magnetic field sweep used for the DNP sequences. It can sweep up to $\pm40$\,mT with a rate of $\sim0.3$\,mT/$\mu$s. After DNP, the magnetic field is raised from 0.36\,T to 0.52\,T and the proton polarization is measured using a pulse NMR system based on a Spincore RadioProcessor card. This is possible thanks to the extremely long proton spin-lattice relaxation time in naphthalene of $>800$\,h \cite{Quan:2019}. The long spin-lattice relaxation time is ideal for testing DNP theories and is one of the reasons for choosing this triplet system.

The relatively short life time of the triplet state requires a repetitive DNP experiment.  Synchronized to the exciting laser pulse, is a strong microwave pulse of 10\,$\mu$s length at fixed frequency, which is applied while the field is adiabatically swept for performing ISE, SSE or ASE. Using a field sweep instead of a frequency sweep has the advantage that the microwave power stays constant over the sweep and is not modulated by the $Q$ of the cavity.

\begin{acknowledgement}

This work was supported by the National Institutes of Health through grants GM132997 and GM132079 to RGG and the Swiss National Science Foundation through a grant to YQ (No.\ P500PN\_202639). We acknowledge the early contributions of Dr. T. V. Can on the ISE and SSE experiments.

\end{acknowledgement}


\bibliography{ISE}

\end{document}